\newcommand{\E}{{\rm e}}

\newcommand{\lt}{\left}
\newcommand{\rt}{\right}

\newcommand{\be}{\begin{equation}}
\newcommand{\ee}{\end{equation}}
\newcommand{\bea}{\begin{eqnarray}}
\newcommand{\eea}{\end{eqnarray}}
\newcommand{\beas}{\begin{eqnarray*}}
\newcommand{\eeas}{\end{eqnarray*}}
\newcommand{\rf}[1]{(\ref{#1})}
\newcommand{\bft}{\begin{figure}[t]}
\newcommand{\bfb}{\begin{figure}[b]}
\newcommand{\bfh}{\begin{figure}[h]}
\newcommand{\ef}{\end{figure}}
\newcommand{\bi}{\begin{itemize}}
\newcommand{\ei}{\end{itemize}}

\documentstyle[aps, pre, epsf]{revtex}
\begin{document}

\title{Near action degeneracy of periodic orbits in systems with non-conventional time reversal}

\author{P.A. Braun, F. Haake, S. Heusler}
\address{Fachbereich Physik, Universit\"at Essen,
45\,117 Essen, Germany }
\date{\today}
\maketitle
\begin{abstract}
€
Recently, Sieber and Richter calculated semiclassically a first off-diagonal contribution to the orthogonal form factor for a billiard on a surface of constant negative curvature by considering orbit pairs having almost the same action. For a generalization of this derivation to systems invariant under non-conventional time reversal symmetry, which also belong to the orthogonal symmetry class, we show in this paper that it is necessary to redefine the configuration space in an appropriate way. 

\end{abstract}€
€
\section{Introduction}

The form factor $K(\tau)$ is defined as the  Fourier transformation 
of the two point energy correlation function of the quantum system where
$\tau$ is time measured in units of the Heisenberg time $T_H$.
If the Gutzwiller trace formula is used for the density
of energy levels the form factor becomes a double sum over the classical periodic orbits $\gamma$ with the period $T=\tau T_H$:
\be
K(\tau)=\lim_{\hbar\to 0}\frac{1}{T_H}\lt\langle\sum_{\gamma,\gamma'}
A_\gamma A^*_{\gamma'}\E^{i(S_\gamma-S_{\gamma'})/\hbar}
\delta\lt(T-\frac{T_\gamma+T_{\gamma'}}{2}\rt)\rt\rangle  \label{sumform} \ \ .
\ee
Here, $A_\gamma$ are the stability coefficients of the orbits; the angle
brackets signify average over a small interval of time \cite{haake}.

Only pairs  of orbits whose action difference is not large compared with $\hbar$ can make a contribution surviving the time averaging. The diagonal approximation
\cite{berry} takes into account the diagonal terms $\gamma=\gamma'$ and pairs of mutually time reversed orbits in the case that the dynamics of the system is invariant with respect to time reversal (TR). The result of the diagonal approximation is then
\be
 K_{\rm{diag}}(\tau)= 2\tau .     \label{diag}
\ee
which is to be compared
with the form factor of the  Gaussian ensemble of random  orthogonal matrices
for $0<\tau<1$,
\be                                                  
K^{GOE}(\tau)=2\tau-\tau\log (1+2\tau).  \label{rnd}
\ee

Obviously, the diagonal approximation reproduces
only the first term of the Taylor expansion of the random matrix form factor.
It may be  expected that taking into account less obvious pairs of
 periodic  orbits with small action difference
 higher terms of  the random matrix form factor
\rf{rnd} will be recovered, in line with the conjecture stated by Bohigas, Giannoni and Schmidt \cite{bohigas}.
In the theory of disordered systems the so-called weak diagonal approximation was used by
 Smith {\it et al} \cite{lerner} who  stressed the importance of
pairs of trajectories with the multiple-loop topology. The  pairs of orbits
whose contribution seems to be responsible for the higher
order terms of the form factor  in the case of clean chaos
have  been discovered only recently
by Sieber and Richter \cite{sieber}.

The Sieber-Richter  pair of orbits is schematically
shown in Fig.1.  One of its members
contains a self-intersection with a  small crossing angle
$\epsilon$  and consists of two loops, one
of which is passed clockwise and another one counterclockwise.
Playing with small deformations of such an orbit
it can be demonstrated that there exists
 a partner  periodic orbit which is almost  everywhere
exponentially close to the original one; however, at one place in configuration space the partner orbit has an avoided crossing at the place of a self-intersection.

\begin{figure}[1]
\begin{center}
\leavevmode
\epsfxsize=0.45\textwidth
\epsffile{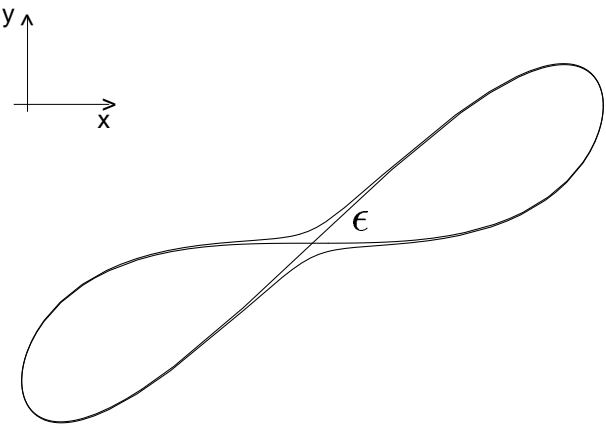}
\end{center}
\caption{Sieber-Richter pairs in configuration space}
\label{1}
\end{figure}

The partner can be regarded as  the same two-loop construction as the  orbit with crossing, however, 
now  both loops are  passed in the same sense of traversal.
The time reversal symmetry is essential
for the existence of the partner; otherwise retracing
the motion along one of the loops would not be allowed.
The difference of actions  of the orbit and its partner
is 
\be
\Delta S\sim \frac{p^2\epsilon^2}{2 m\lambda}
\ee
where $\lambda$ is the Lyapunov constant of the orbit. For small $\epsilon $
it can be of the order of $\hbar$.

The stumbling block is to evaluate the
number of the Sieber-Richter  pairs and to sum over their contributions.
Up to now it has been done only for the billiards in the space of constant
negative curvature where all orbits have the same Lyapunov constant and the
Maslov index equals zero.   Contribution of all  pairs  Fig.1
have been found to be $-2\tau^2$   \cite{sieber}
which coincides with the term of the order $\tau^2$ in the expansion
of the random matrix form factor \rf{rnd}.

One of the open problems in the Sieber-Richter theory is connected
with the so-called
non-conventional time reversal (NCTR) symmetry. It is often encountered in systems
in external magnetic field provided that there exists a suitable spatial
symmetry \cite{haake}. Their Hamiltonian is invariant under the conventional time reversal combined with an appropriate  spatial symmetry operation. Each periodic orbit of a
NCTR-symmetric system has a twin with the same action. However, unlike
the  conventional TR dynamics, the twin
is not the same orbit retraced backwards.

The statistical properties of the energy levels of
the systems with NCTR and the conventional TR symmetry are identical.
In particular, the form factor of the systems with NCTR is given by
\rf{rnd}. Therefore, the diagonal approximation is not sufficient, and
pairs of orbits with very close action must also exist and contribute
to $K(\tau)$.

At the first glance it seems that the Sieber-Richter arguments collapse
in the case of  NCTR.     Even  a very weak magnetic field
destroys the closeness of  action in the pair  because of the totally
different magnetic flux (due to the fact that one of the loops is passed
in the opposite sense by the members of the pair).
In a stronger field whose impact on the trajectory cannot
be neglected the pairs  in Fig. 1 simply
cannot exist because passing the same loop in the opposite direction would
contradict the equations of motion. The  spatial symmetry
implied by NCTR does not help.  Since  the  correct semiclassical explanation
of the form factor \rf{rnd} must be essentially the same, be it systems with TR or NCTR, the
inability to produce the contributing pairs of
orbits  in the  NCTR-symmetric systems might compromise  the
whole Sieber-Richter theory. Below we show that these apprehensions are groundless, and the
case of NCTR does not present any new difficulties.

Consider a two dimensional motion in the orthogonal uniform
magnetic field ${\bf B}=B{\bf e}_z$ and suppose that the
potential energy has the symmetry
\be
V(x,-y)=V(x,y) \ .
\ee
We shall use  the gauge $A_x=-By,\;A_y=A_z=0$.
Then the classical Hamiltonian of the system
\be
H=\frac{1}{2m}\lt(p_x+\frac{eBy}{c}\rt)^2+\frac{p_y^2}{2m}+V(x,y)
\ee
will be invariant with respect to NCTR
consisting of the conventional time reversal TR (changing the
sign of the canonical momenta) followed by the reflection
in the $x$ axis of the plane (replacement $y\to -y,\;p_y\to -p_y$).
An arbitrary periodic orbit of our system with the  trajectory
\be
x=x(t),\quad y=y(t),\label{orb}
\ee
and the momentum $p_y(t)=m \dot{y}(t)$ along $y$ has a NCTR twin,
\be
\tilde{x}(t)=x(-t), \quad\tilde{y}(t)=-y(-t),
\quad\tilde{p}_y(t)=p_y(-t) .
\ee
The  twin has the same magnetic flux and action.  Its trajectory
is obtained from the original one by reflection in the $x$ axis while the
sense of traversal on both orbits is the same. Let us canonically transform the variables as
\be
x'=x, \;y'=p_y,\quad  p_x'=p_x,\;p_y'=-y
\ee
and consider how our two orbits project on
the new configuration
space $x'y'$. The original orbit  will be described by the equations
\be
x'(t)=x(t),\quad y'(t)=p_y(t)    \label{orb'}
\ee
with $x,p_y$ given by Eqs. \rf{orb}.
Its NCTR twin obeys
\be
\tilde{x}'(t)=x(-t),\quad \tilde{y}'(t) =p_y(-t)   \ \ . \label{NCTRorb'}
\ee
The only difference between \rf{orb'} and \rf{NCTRorb'} is the change
of sign of $t$. Consequently, in the $xp_y$ plane the two orbits are
depicted by the same closed curve traversed in opposite directions:
NCTR acts on a periodic orbit in the coordinates $xp_y$  exactly
like the usual time reversal in the ordinary configuration space.

Now it is easy to see that in the case of NCTR
the Sieber-Richter arguments for the
existence of the two-loop pairs of orbits Fig.1 remain fully valid. 
However, these pairs may only exist in the $xp_y$ plane (Fig.2a) which is the configuration 
plane in the new coordinates. Only in this
projection of the phase space to a two-dimensional submanifold switching the sense of traversal of a loop caused by the
replacement of crossing by an avoided crossing is compatible with the
equations of motion. Therefore, only in the $xp_y$ plane pairs as depicted in 
Fig.1 exist and have close actions in strong magnetic fields.
Evaluation of the number of Sieber-Richter pairs for the system with NCTR and their contribution to
the form factor therefore does not lead to new difficulties as compared to the problem of conventional TR-invariant systems.

\begin{figure}[2]
\begin{center}
\leavevmode
\epsfxsize=0.45\textwidth
\epsffile{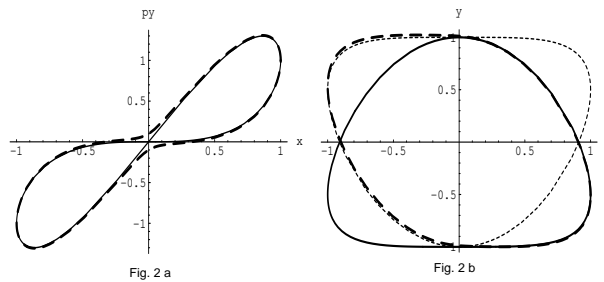}
\end{center}
\caption{Sieber-Richter pairs in a NCTR system}
\label{fig2}
\end{figure}

It is instructive to see how the pairs Fig. 2a look like when the respective
motion is  projected onto the usual configuration space $xy$ (Fig.2b). They have little in common 
with the double-loop Sieber-Richter€pairs of Fig.1.
However, the general idea of building
a new orbit with practically the same action by gently reconnecting parts of
the original orbit and its NCTR twin is still obvious. Depending on the projection chosen, the criterion for finding 
Sieber-Richter pairs changes. In  systems with NCTR one may either look for the 
two loop orbits with small opening angle in the $xp_y$ plane or search
for the orbits like in Figure 2b. The latter may be preferable for systems like billiards in the magnetic
field whose  trajectory in the $xp_y$ plane is discontinuous.

It may be somewhat puzzling that a jump in representation is needed for
the Sieber-Richter treatment when the magnetic field is switched on:
instead of $xy$ space we must shift to $xp_y$. However such a jump is
only natural in view of  the change of the universality class
of the dynamics. A chaotic system with the NCTR
symmetry belongs to the Gaussian orthogonal ensemble only in the
presence of the magnetic field.  When the field
is switched off the spectrum splits
into two independent subspectra (even and odd with respect to 
$y\to-y$).  Superposition of  two such spectra whose
levels may cross each other creates a specific ensemble obviously
different from GOE, usually called GOE$\times$GOE.

The systems with the plane of symmetry in the uniform magnetic field
constitute the most important but not the only example of NCTR.  Consider
e.g. a two-dimensional system with a center of symmetry
\be
V(-x,-y)=V(x,y)
\ee
in an extremely  non-uniform  magnetic field
\be
{\bf B}=y {\bf e}_z \ \ .
\ee
In the  gauge
$
A_x=-By^2/2 ,\;A_y=A_z=0
$
 the Hamiltonian of the system is invariant with respect to the NCTR composed
of the time and spatial inversion which means that the momenta are unchanged.
The partner obtained from a periodic orbit by this symmetry operation
coincides with the original orbit if we draw it in the plane of
momenta $p_xp_y$, hence it is in this plane that the
Sieber-Richter pairs are described by small intersection angle $\epsilon$.

To summarize, the Sieber-Richter double loop pairs  Fig. 1
or Fig. 2a may be observed only in the configuration space whose
generalized coordinates are unchanged by the particular NCTR.  A
periodic orbit and its NCTR twin are depicted in this space by the same curve
traversed in opposite direction. This may be the usual coordinate space
$xy$ in systems with conventional time reversal symmetry, the $xp_y$ space in
systems with the NCTR $t\to-t,y\to-y$, and the $p_xp_y$ space if NCTR is
described by $t\to -t,x\to -x,y\to -y$.  An attempt to break out of
this symmetry-dictated space by a canonical transformation
mixing the coordinates and momenta will immediately strip the Sieber-Richter
pairs of their double-loop, intersection/avoided-crossing appearance. 
However, using appropriate citeria to describe the close action partners can be recognized in principle in any two dimensional projection of the phase space.

\vspace{2cm}

 \appendix{{\bf Appendix}}

Calculations in \cite{sieber} are based on the geometric evaluation
of the shifts of momenta  using the fact that
velocities and momenta in the conventional
coordinate space are practically identical.
After our canonical transform  the
connection between the new ``velocities'' and the  momenta  becomes
more complicated. In particular the direction of the momentum is no
longer tangent to the trajectory in the new configuration space. That
means that the Sieber-Richter result has to be rederived.

 We assume the existence of a   self-crossing periodic orbit with a
small opening angle like the one shown in Fig. 1, 2a.
  A coordinate frame is introduced with its  origin at the point of
 crossing, and  the $x'$ axis  along the bisector of the small angle.
Consider the Poincar\'{e} section  at $x'=0$ with the
coordinates  $y'$ and $p_y'$ on the crossing plane. The true
Poincar\'{e} map (Fig.3) is obtained when
passages of the $x'=0$
 plane with a certain sign of $\dot{x'}$, say, $\dot{x'}>0$ are marked. The
self-crossing orbit will then be depicted by a periodic point   $O$ on
 the $p_y'$ axis with $p_y'$ positive and small.  The self-crossing
TR (NCTR) twin of
this orbit will produce another periodic point $O'$ symmetrical with
respect to the $y'$ axis: $y'=0,p_y'<0$.

\begin{figure}[3]
\begin{center}
\leavevmode
\epsfxsize=0.45\textwidth
\epsffile{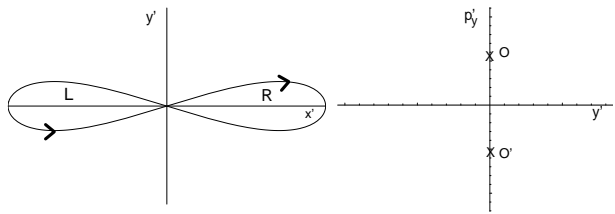}
\end{center}
\caption{Poincar\'e map of the Sieber-Richter Orbit}
\label{fig3}
\end{figure}

We shall concentrate, however, on the submaps $R$, $L$ 
of the Poincar\'{e} map describing the transform of $y',p_y'$  generated by the right and
left loop of the orbit.
We shall mark the crossing of the $x'=0$ plane both for
$\dot{x'}>0$
 and $\dot{x'}<0$; such a break of the
rules is needed since $R$ and $L'$ are not true Poincar\'{e} maps.
We shall also be interested
in the TR (NCTR) submaps obtained by passing the loops of the orbit in the
direction opposite to Fig. 3; these  will be denoted $R'$ and $L$, 
respectively. The periodic point $O$ of the total Poincar\'{e} map is
simultaneously the periodic point of the submaps $R$ and $L$ whereas
$O'$ is the periodic point of the submaps $L'$  and $R'$.

\begin{figure}[4]
\begin{center}
\leavevmode
\epsfxsize=0.45\textwidth
\epsffile{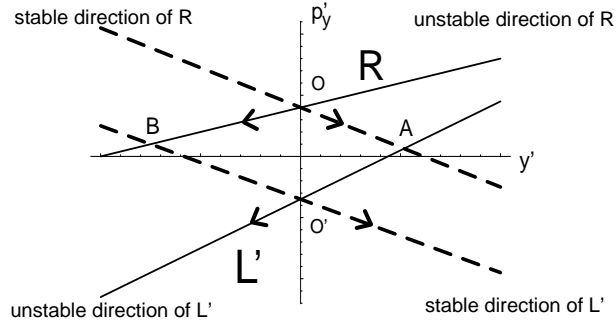}
\end{center}
\caption{Stable and instable manifolds of the right submap R and the time-reversed submap L'. A point P in the vicinity of A will be mapped to a point Q in the vicinity of point B. Fine-tuning of the initial point P leads to a periodic orbit with avoided self-crossing.}
\label{fig4}
\end{figure}

Each submap can be characterized by its  stability matrix
connecting the initial and final deflections of the $y'$ coordinate and
momentum from the  periodic point of the respective submap.
 We shall need $M_R$ (the stability matrix of the right loop passed as
it is shown in Fig. 3) and $M_{L'}$ (the one for the left loop followed
in the direction opposite to  Fig.3).
 The  large eigenvalues $\Lambda_R$ and $ \Lambda_{L'}$ of these two matrices
can be evaluated as $\sim \exp (\lambda T)$ where $\lambda$ is the
 Lyapunov constant and $T=T_R,T_{L'}$ is the period of the respective
loop. The periods of the orbits in the sum for the
form factor  \rf{sumform} are of the order
  $\hbar^{-1}\to\infty$, therefore
the larger eigenvalues of the stability matrices are exponentially large
whereas the smaller ones ($1/\Lambda$) are exponentially close to zero.
  The respective eigenvectors
determine the unstable  and stable directions of each of the submaps.

Consider Fig.4 where periodic points $O,O'$ of the submaps $R,\;L'$ and their
stable and unstable directions are shown. Let us investigate the
application of the submap $R$ to an initial point $P$ chosen
in the vicinity of the crossing $A$ of the stable direction of $R$ and
unstable direction of $L'$. We shall represent the initial
radius-vector by an expansion in powers of $\Lambda_R^{-1}$,
\be
{\bf r}_P= {\bf e}_s^R\lt(l_{OA}+\frac{c_1}{\Lambda_R}+\ldots\rt) +
{\bf e}_u^R \lt(\frac{l_{OB}}{\Lambda_R}+\frac{d_2}{\Lambda_R^2}+\ldots
\rt) \ \ .
\ee
 Here     ${\bf e}_s^R, {\bf e}_u^R $ are the eigenvectors of the stability
matrix of $R$ along
the stable and unstable directions, $l_{OA}$ and  $l_{OB}$ are
distances from $A$ and $B$ to the periodic point $O$ ($B$ is
the crossing of the unstable direction of $R$ and stable
direction of $L'$); the coefficients
$c_1,d_2,\ldots$ are so far undetermined.

After the loop $R$ has been completed the point $P$ will be  mapped to the
point $Q$ obtained by squeezing along the stable and stretching along
the unstable direction with the coefficient $\Lambda_R$:
\be
{\bf r}_Q={\bf e}_s^R\lt(\frac{l_{OA}}{\Lambda_R}
+\frac{c_1}{\Lambda_R^2}+\ldots\rt) +
{\bf e}_u^R \lt(l_{OB}+\frac{d_2}{\Lambda_R}+\ldots
\rt) \ \ .
\ee
It is seen that $Q$ is infinitely close to the  crossing point $B$
of the unstable direction of $R$ and the stable direction of $L'$;
 thus the distance of $Q$ from the stable manifold of $L'$ is
exponentially small and depends on  the coefficients $c_1,d_2,\ldots$.

Now consider the loop $L'$ taking $Q$ as its initial point.
The loop will practically annihilate the stable component
${\bf e}_u^R l_{OB}$
  and  place
the final point  somewhere on its unstable manifold. The exact
position of the final point on the unstable manifold of $L'$  depends
on
$c_1,d_2,\ldots$; these  can be  finetuned  so that the final point will
coincide with the initial point $P$ of the loop $R$.
\footnote{More accurately, demanding
that the final point of the second loop coincides with the
initial point $P$ of the first loop we obtain a set of equations for
consecutive definition of $c_1,d_2,c_2,d_3\ldots$. The coefficients in the
expansion will not be growing and convergence for ${\bf r}_P$ guaranteed
 provided $\Lambda_R<\Lambda_L$ which
can be assumed wihout loss of generality.}
But that would mean that $P$ is a periodic point of the
composition  of the  submaps $R$ and $L'$.
It corresponds to a new periodic orbit  composed of the
deformed loops $R$ and $L'$. The new orbit  crosses the ``true'' Poincar\'{e}
map at the point  infinitely close to the crossing of the
stable manifold of $R$ and unstable manifold of $L'$. The TR (NCTR) twin
of the new
periodic orbit can be found by considering the sequence of the submaps
$R'$ and $L$.    The new orbit and its twin are of course the
 Sieber-Richter
partners with avoided crossing of the orbit with self-intersection.

As can be seen we have not used
  the  connection between the
velocities and momenta of the usual configuration space.
Note also that the effect of the pair formation  seems to remain
 in force even
when the nonlinear corrections to the submaps $R,L'$ are taken into account,
 with their stable and unstable manifolds depicted by curves
rather than straight lines, as long as the stable and unstable manifold intersect only once. For small angle 
$\epsilon$, this condition should be fulfilled.

\end{document}